\documentclass{ws-ijmpd}
\usepackage[super,compress]{cite}
\usepackage{graphicx}
\usepackage{amsfonts}
\usepackage{empheq,amsmath}
\usepackage{amssymb}
\usepackage{hyperref}
\usepackage{breakurl}
\usepackage[latin1]{inputenc}
\usepackage{mathrsfs}
\begin{document}
\markboth{F. P. Poulis \& J. M. Salim}
{Weyl Geometry and Gauge-Invariant Gravitation}

%
\catchline{}{}{}{}{}
%

\title{WEYL GEOMETRY AND GAUGE-INVARIANT GRAVITATION}

\author{F. P. POULIS}

\address{Universidade Federal do Maranh\~{a}o / CCCT\\Av. dos Portugueses, 1.966. S\~{a}o Lu\'{\i}s, MA, Brazil. CEP 65080-805.\\
fp.poulis@ufma.br}

\author{J. M. SALIM}

\address{Centro Brasileiro de Pesquisas F\'{\i}sicas / ICRA\\Rua Xavier Sigaud, 150. Rio de Janeiro, RJ, Brazil. CEP 22290-180.\\
jsalim@cbpf.br}

\maketitle

\begin{history}
\received{Day Month Year}
\revised{Day Month Year}
\end{history}

\begin{abstract}
We provide a gauge-invariant theory of gravitation in the context of Weyl Integrable Space-Times. After making a brief review of the theory's postulates, we carefully define the observers' proper-time and point out its relation with space-time description. As a consequence of this relation and the theory's gauge symmetry we recover all predictions of General Relativity. This feature is made even clearer by a new exact solution we provide which reveals the importance of a well defined proper-time. The thermodynamical description of the source fields is given and we observe that each of the geometric fields have a certain physical significance, despite the gauge-invariance. This is shown by two examples, where one of them consists of a new cosmological constant solution. Our conclusions highlight the intimate relation among test particles trajectories, proper-time and space-time description which can also be applied in any other situation, whether or not it recovers General Relativity results and also in the absence of a gauge symmetry.

\end{abstract}

\ccode{PACS numbers: 04.20.-q; 04.20.Cv; 04.20.Jb; 04.90.+e}


\section{Introduction}
Special Relativity requires a suitable device in order to have its observables properly measured. The standard, and only one, measuring tool in accordance with their postulates is the so called \emph{light-clock}\cite{Blokhintsev}. When moving to a curved space-time, provided with local Lorentz invariance, the light-clock still remains as the only appropriate one to make measurements, in particular, those regarding the geometrical objects of the theory\cite{Graves}.

From the use of such device, Ehlers, Pirani and Schild (EPS), corroborated by Woodhouse, have shown\cite{EPS,Woodhouse} that one is able to consider a more general geometry than the pseudo-riemannian\footnote{Henceforth we will call it just \textit{riemannian} to shorten.} to describe space-time, namely, a Weyl geometry\cite{Weyl}. This one generalizes Riemann geometry by allowing vector moduli to vary along the manifold when they are parallel propagated. To provide this, it is introduced a new geometrical object ($\omega_\mu$) responsible for this variation which has no association at all with any other physical field nor the metric of the manifold. In other words, Ehlers, Pirani and Schild conclusion is that space-time can be geometrically described in a more general way than it was originally conceived, by associating two geometrical objets to it, instead of only the metric as in General Relativity (GR).

Our intent with this paper is then to investigate to what extent the consideration of this new geometrical object responsible for moduli variation of parallel transported vectors can affect GR predictions.

In the past decades, many work has been done in this direction\footnote{See Refs.~\citen{Novello2,Novello3,Novello1,Novello4,Salim-Sautu1,Salim-Sautu2,Salim-Sautu3,Salim-Sautu4,Salim-Sautu5,Scholz:2011za} and references therein.} and the main geometrical aspects o Weyl geometries has already been extensively pointed out, specially concerning its gauge invariance. Furthermore, when dealing with the particular case where $\omega_\mu$ is irrotational, it is also widely emphasized the equivalence between this and Riemann geometry, since the later would only correspond to a specific gauge.

However, depending on the way other physical fields couples with geometry, the gauge invariance can be broken and there will be no equivalence at all with riemannian geometry nor GR.

In this paper, the main issues will be reviewed and matter fields will be coupled to geometry in a way that preserves its gauge symmetry. Therefore, our whole formalism will seem to be irrelevant, since GR would represent a specific gauge and, thus, we do not provide any new prediction. Nevertheless, it should be noted that the resulting gauge-invariant theory of gravitation is already very interesting by itself. As it will be shown, the gauge transformation involves a conformal transformation of the metric, and then we believe a step has been given in the direction of a conformal invariant theory of gravitation, which would have some deep theoretical implications.\cite{Bekenstein-Conformal-invariance}

Moreover, despite the frustrating lack of new observational consequences, some great contributions come from the careful study of the theory, which allows us to define the observables in a gauge independent way rather then evoking that specific one which reduces the theory to GR. By doing this, we are also able to consider those situations where the gauge symmetry is broken (and hence there is no equivalence to GR at all) and still provide a proper interpretation of observables in terms of the geometrical objects. We do not work with this possibility in this paper, but we will anticipate the consequence of breaking the gauge symmetry by introducing a lagrangian term for $\omega_\mu$ in the gravitational action.

Before working on Weyl geometries directly, we will briefly review the geometrical description of gravitation in Sec.~\ref{sec: geometrical formulation}, pointing out some of its main issues. In particular, we will make a precise definition of proper-time by imposing that it must ensure the postulate for freely falling test particles trajectories. Additionally, it will be emphasized the intimate relation between space-time description and the observers' proper-time measurements. Then, we will point out the close relationship among free test particles trajectories, their corresponding action, proper-time and space-time description.

Next, we introduce Weyl geometry as a consequence of EPS axiomatic approach. A departure from GR in the expression for the connection coefficients is obtained and a gauge symmetry in the geometry description is noted. It is then required that this same symmetry should be present in the observers' description of space-time, which is naturally verified from the proper-time definition we provide.

In Sec.~\ref{sec: variational formulation} we will be concerned with a lagrangian formulation for the theory. Since it is not known such kind of formulation for the most general case of Weyl geometries, we will restrict ourselves to its particular case called Weyl Integrable Space-Time (WIST), in which $\omega_\mu = \partial_\mu\omega$ ($\partial_\mu \equiv \partial / \partial x^\mu$) and moduli variation is integrable on the manifold. Furthermore, according to Ref.~\citen{Jurgen}, only in this case we can reconcile the theory with others well established formalisms in physics, like the Hamilton-Jacobi one.

In WIST, Riemmann geometry is recovered by a simple gauge transformation. Hence, both of them must be equivalent, since the later is just a specific gauge of the former. Therefore, although the difference from GR in the connection coefficients might seem to affect freely falling particles trajectories, they are described by observers precisely the same way as they are in GR (or any other gauge) when one adopts a proper-time in consistency with geometry's gauge symmetry.

This equivalence was already realized a long time ago, as well as the recognition that only gauge-invariant quantities can be physically meaningful. However, this criterion of gauge-invariance has never been provided with such a conceptual support as in our approach, where it naturally arises and relies in the very fundamental principles of the theory. The whole point is the realization that a space-time described by the proper-time we have defined gives rise to a gauge-invariant tensor that plays the same role as the metric in GR,\footnote{Naturally, when recovering the riemannian geometry by setting $\omega_\mu = 0$, this tensor will be the metric itself.} which we will call \emph{effective metric}. Therefore, it is just a matter of working with this gauge-invariant tensor in the same way it is done with the metric in GR; this procedure will be made evident and justified in this paper.

We present, then, actions for both the space-time geometry and free test particles trajectories consistently with their postulates and gauge symmetry. It is also established a coupling with other physical fields in a way that the whole theory presents the same gauge symmetry of the geometry and we end up then with a gauge-invariant gravitation theory which is equivalent to GR. At this point, the theory will look very similar to a conformal transformation in GR and we will try to elucidate this is not the case.

The effective metric consists of a scalar factor multiplying the metric, nevertheless, the former does not arise from a conformal transformation in the later but rather from the consistent proper-time definition we made. If this would not result in such an expression for the effective metric, there might not be any resemblance at all with a conformal transformation.

Afterwards, in Sec.~\ref{sec: vacuum solution}, we solve the equations for a static and spherically symmetric vacuum, whose solution constitutes an explicit example of how a consistent space-time description actually gives the same results of GR in any gauge, reinforcing the need of a space-time description by means of the effective metric.

In Sec.~\ref{sec: thermodynamics}, we perform a description of the source fields according to their thermodynamical regime in this new gauge-invariant context and in Sec.~\ref{sec: examples} we argue that each of the geometric fields possesses a true physical reality, although their specific roles are rather vague by virtue of the gauge symmetry. Nevertheless, it will be given two specific examples of geometries that differ only in $\omega$ and are associated with totally different geometries and, from the results of Sec.~\ref{sec: thermodynamics}, we will determine to which physical system they correspond.

We conclude that this gauge-invariant theory is theoretically more interesting and consistent than GR since it considers a more general geometry than the riemannian and meets the conclusion of EPS axiomatic approach to describe space-time, besides the gauge invariance which is of great theoretical interest by itself. Furthermore, due to the specific coupling with matter fields we chose, we obtained a gravitation theory that gives the same predictions of GR. The only restriction made was for the sake of a variational formulation and happened to have the fortunate property of recovering Riemann geometry by a gauge transformation. Thanks to this property, it can be easily verified that the interpretation given for the observables in terms of the geometrical objects is indeed consistent.

Although there is not any new observational consequence when compared to GR, we believe our conclusions are particularly important to illustrate the theory main ideas, specially concerning the relation among test particles lagrangian, proper-time and space-time description, characterizing the corresponding frame. From this, we consider that our conclusions may perfectly well be extended to other cases that envolve a non-riemannian connection and/or a different test particle action as well, regardless the presence of any gauge symmetry.

\section{Geometrical Formulation of Space-Time} \label{sec: geometrical formulation}
Space-time is considered as a local Lorentz invariant manifold endowed with a \emph{metric tensor} ($g_{\mu\nu}$) which gives the \emph{interval}:

\begin{equation}
ds^2 = g_{\mu\nu}dx^\mu dx^\nu \; .
\label{interval definition}
\end{equation}

\noindent The metric tensor is symmetric in their indices, has $g^{\mu\nu}$ as inverse ($g^{\mu\lambda}g_{\lambda\nu} = \delta^\mu_\nu$) and those are the ones used to lower or raise indices.

Besides the metric, there is also another geometrical object called \emph{connection} ($\Gamma^\alpha_{\mu\nu}$) which enters in the expression for the change in an arbitrary vector due to the geometry of the manifold. That is, for any vector field $A^\mu$ we have for an infinitesimal displacement $dx^\alpha$ between two nearby points the change due to the curvature given by:

\begin{equation}
\delta A^\mu = -\Gamma^\mu_{\alpha\beta}A^\alpha dx^\beta \; .
\label{geometrical variation}
\end{equation}

\noindent It is important to emphasize that this is the \emph{total} change in the vector which comes \emph{exclusively} from the geometry of the manifold.

From this definition we have the following difference in the vector field apart from the geometrical contribution:

\begin{equation}
DA^\mu = dA^\mu - \delta A^\mu \; ,
\label{difference apart from geometry}
\end{equation}

\noindent where $dA^\mu = \partial_\nu A^\mu dx^\nu$ is the total difference in the vector field along the same displacement. We see that, when $DA^\mu = 0$, the only changes in the vector field are those provided by geometry.

If the displacement is along a curve parameterized by $\sigma$, we have from (\ref{geometrical variation}) and (\ref{difference apart from geometry}):

\begin{equation}
DA^\mu = \left(\partial_\nu A^\mu + \Gamma^\mu_{\alpha\nu}A^\alpha\right)\frac{dx^\nu}{d\sigma}d\sigma = \left(u^\nu\nabla_\nu A^\mu\right)d\sigma \; ,
\label{difference in terms of covariant derivative}
\end{equation}

\noindent where we have the \emph{tangent} to the curve and the \emph{covariant derivative} of a vector $A^\mu$ defined respectively by:

\begin{gather}
u^\nu \equiv \frac{dx^\nu}{d\sigma} \; , \label{four-velocity definition} \\
\nabla_\nu A^\mu \equiv \partial_\nu A^\mu + \Gamma^\mu_{\alpha\nu}A^\alpha \; . \label{covariant derivative definition}
\end{gather}

\noindent Once again, if $u^\nu\nabla_\nu A^\mu = 0$, it means that all changes in $A^\mu$ come exclusively from the curvature of the manifold. In other words, if this quantity is different from zero, then the total change in the vector field cannot be attributed to the geometry; there are extra contributions besides the geometrical ones.

Now, we postulate that free test particles have their space-time trajectories entirely determined by geometry, therefore we must have their tangent vector, which is the particle's four-velocity now, satisfying:

\begin{equation}
u^\nu\nabla_\nu u^\mu = 0 \; .
\label{geodesic equation for proper time}
\end{equation}

\noindent This equation cannot be satisfied by any parameter. It defines, instead, a specific one (up to affine transformations) which we will call \emph{proper-time}, $\tau$.

It is interesting to note that the same curve it gives, called \emph{geodesic}, could be obtained from the more general equation:

\begin{equation}
u^\nu\nabla_\nu u^\mu = f\,u^\mu \; ,
\label{geodesic equation for a general parameter}
\end{equation}

\noindent where $f$ is some function of the coordinates. In fact, the only difference from the previous case is that the tangent vector no longer changes only according to the geometry, but has some extra variations. However, these are along the tangent's direction and so it will not change the curve, but rather change the tangent vector modulus in each point. But, since this modulus is related to its parametrization, we see that imposing a trajectory solely determined by geometry is the same as setting a specific parameter along the curve.

As it has been already pointed out by Perlick,\cite{Perlick} this relationship between free test particle trajectory and proper-time is indeed completely analogous to Newton's first law, which states that a free moving body not only follows a straight line but it also has a constant velocity along the path, defining thus the observer's clock and rulers properly. Eq.~(\ref{geodesic equation for proper time}) can be viewed then as a generalization of this law for a curved space-time, setting the observer's proper-time correctly.

Eq.~(\ref{geodesic equation for a general parameter}), however, will be very useful later on when we will be concerned in obtaining the proper-time, but we need to know some other features of the geometry considered to solve for this parameter.

\subsection{Axiomatic approach and Weyl geometry} \label{subsec: axiomatic approach}
Assuming that observers make use of a measuring device based on light rays emission and reception (light-clock) and postulating some properties of freely falling test particles and light rays propagation, Ehlers, Pirani and Schild \cite{EPS} concluded that light-clocks allow one to consider Weyl geometry as the most general one to describe space-time. That is, if space-time geometry is to be measured by a light-clock, which is the most appropriate for this purpose \cite{Graves}, then Riemann geometry represents a restriction among the possibilities this measuring tool admits; one should consider the one developed by Weyl instead.

Before introducing this kind of geometry, let us briefly describe the light-clock. It consists of two mirrors with fixed distance between them which reflect a light ray one to another repeatedly. Each reflection can be used as the unit of proper-time lapse by the observer who carries the light-clock. To measure the space-time distance, $S$, to a nearby event, the observer sends a light ray at proper-time $\tau_1$ which is reflected at the event and arrives back to the observer at proper-time $\tau_2$. The space-time distance is given in terms of those measurements by \cite{Blokhintsev} $S = \tau_1 \tau_2$. From this, we can clearly see it is according to its own proper-time that observers describe space-time.

This is a crucial issue, because proper-time is defined as the one for which the function $f$ in (\ref{geodesic equation for a general parameter}) is zero. However different geometries may provide different proper-times and, therefore, space-time will not be described by the interval (\ref{interval definition}) in general.

Differently from Riemann geometry, in Weyl it is also possible that the modulus of an arbitrary vector $A^\mu$ changes when parallel propagated, \textit{i.e.}, when only geometry acts on the vector, besides the change in its direction. This modulus varies according to the equation:

\begin{equation}
dl^2 = l^2\omega_\alpha dx^\alpha \; ,
\label{modulus variation}
\end{equation}

\noindent where $l^2 = g_{\mu\nu}A^\mu A^\nu$ and $\omega_\alpha$ is a new geometric vector responsible for moduli variation. In the case this vector is zero we must naturally recover Riemann geometry.

One can easily see that this equation is invariant under the following gauge transformation:

\begin{equation}
\left\{\begin{array}{rcl}
g_{\mu\nu} & \rightarrow & \bar{g}_{\mu\nu} = e^{\Lambda}g_{\mu\nu} \; , \\
\omega_\mu & \rightarrow & \bar{\omega}_\mu = \omega_\mu + \partial_\mu\Lambda \; ,
\end{array}
\right.
\label{gauge transformation for general Weyl}
\end{equation}

\noindent where $\Lambda = \Lambda(x^\mu)$ is an arbitrary function of the coordinates. Although this invariance regards modulus variation, the whole geometry description will be symmetric under this transformation.

Since Eq.~(\ref{modulus variation}) holds for any $A^\mu$ and trajectory, we can obtain that

\begin{equation}
\nabla_\gamma g_{\alpha\beta} = \omega_\gamma g_{\alpha\beta} \qquad \Leftrightarrow \qquad \nabla_\gamma g^{\alpha\beta} = -\omega_\gamma g^{\alpha\beta} \; .
\label{Weyl metricity}
\end{equation}

\noindent This equation, in turn, can be solved for the connection to give:

\begin{equation}
\Gamma^{\alpha}_{\mu\nu} = \frac{1}{2}g^{\alpha\beta}\left(\partial_\mu g_{\beta\nu} + \partial_\nu g_{\mu\beta} - \partial_\beta g_{\mu\nu}\right) - \frac{1}{2}\left(\omega_{\mu}\delta_{\nu}^{\alpha} + \omega_{\nu}\delta_{\mu}^{\alpha} - g_{\mu\nu}\omega^{\alpha}\right) \; .
\label{Weyl connection}
\end{equation}


As it was already stated, the connection accounts for the geometrical effects on vectors, providing the changes they suffer when parallel propagated, according to Eq.~(\ref{geometrical variation}). Moreover, it can be verified that this connection is also invariant under transformation (\ref{gauge transformation for general Weyl}). Therefore, we see that this gauge symmetry is a feature of the geometry that describes space-time and thus the description of it by observers must be compatible with this property, \textit{i.e.}, every measurement involving space-time should be invariant under this gauge transformation as a matter of consistency.

For an arbitrary vector $A^\alpha$, with the help of (\ref{Weyl metricity}) it can be shown that:

\begin{gather}
A_\alpha\nabla_\beta A^\alpha = \frac{1}{2}\partial_\beta\left( A_\alpha A^\alpha \right) - \frac{1}{2}A^\alpha A_\alpha\omega_\beta \; , \label{contraction of A with its derivative I} \\
A^\alpha\nabla_\beta A_ \alpha = \frac{1}{2}\partial_\beta\left( A_\alpha A^\alpha \right) + \frac{1}{2}A^\alpha A_\alpha\omega_\beta \; .
\label{contraction of A with its derivative II}
\end{gather}

\noindent With those expressions in hand, we can solve (\ref{geodesic equation for a general parameter}) for $f$. Contracting this equation with $u_\mu$ and making $u^2 = u_\mu u^\mu$, according to (\ref{contraction of A with its derivative I}) we have:

\begin{gather}
u^\nu u_\mu\nabla_\nu u^\mu = u^\nu \left(\frac{1}{2}\partial_\nu u^2 - \frac{1}{2}u^2\omega_\nu\right) = f u^2 \label{velocity scalar acceleration} \\
\therefore \quad f = \frac{1}{2u^2}\left(u^\nu\partial_\nu u^2 - u^2 u^\nu \omega_\nu \right) \; . \label{solution for f}
\end{gather}

For a geodesic parameterized by $\sigma$, it follows that:

\begin{equation}
u^\nu\partial_\nu = \frac{dx^\nu}{d\sigma}\frac{\partial}{\partial x^\nu} = \frac{d}{d\sigma} \quad \therefore \quad u^\nu\nabla_\nu u^\mu = \frac{1}{2u^2}\left(\frac{du^2}{d\sigma} - u^2 u^\nu \omega_\nu \right)u^\mu \; . \label{geodesic equation in Weyl}
\end{equation}


\noindent This is the most general geodesic equation and holds for any parameter $\sigma$.

Now we are able to determine the proper-time as the one for which the expression between parenthesis in (\ref{solution for f}) or (\ref{geodesic equation in Weyl}) is zero:

\begin{equation}
\frac{du^2}{d\tau} = u^2 u^\nu \omega_\nu \; .
\label{proper-time equation}
\end{equation}

\noindent From this and the first equality in (\ref{velocity scalar acceleration}), we see that $u^\nu u_\mu\nabla_\nu u^\mu = 0$ even for an accelerated particle, which agrees with the (rather mathematical) proper-time definition found in Ref.~\citen{Perlick}.

We can clearly see the solution for proper-time in Weyl will be different from the riemannian case, given by the (gauge-dependent) interval (\ref{interval definition}). Moreover, by performing transformation (\ref{gauge transformation for general Weyl}) we see this equation does not change, which implies that this proper-time solution is also gauge-invariant, as expected, and correctly provides a space-time description in accordance with its geometric properties.\footnote{Notice we have not ascribed any specific form for it; we only demand compatibility with (\ref{geodesic equation for proper time}).}

Furthermore, due to the gauge transformation, the metric alone is completely meaningless and is not sufficient to describe space-time. This means that Eq.~(\ref{interval definition}) cannot provide any interpretation of the coordinates considered. The same role the interval plays in GR is now accomplished by the proper-time solution of the above equation and this should be the one considered to interpret the coordinates. This point will be made very clear with the vacuum solution of Sec.~\ref{sec: vacuum solution}.

Actually, Eq.~(\ref{proper-time equation}) could be written equally well in the form $du^2 = u^2\omega_\nu dx^\nu$, which is precisely the same as (\ref{modulus variation}) for the four-velocity vector. However, although modulus variation is a geometrical feature and holds independently of the parameter chosen to describe the curve, one should notice there is still a dependence on it in the expression for $u^2$, which makes the equation for this case be satisfied by a specific parameter. It is remarkable that when the vector considered in (\ref{modulus variation}) is precisely the tangent to the curve where the displacement takes place the parameter along it is fixed in consistency with (\ref{geodesic equation for proper time}).

In fact, as it was already stated before (\ref{modulus variation}), this equation holds only for a parallel propagated vector and, therefore, Eq.~(\ref{geodesic equation in Weyl}) reveals a striking compatibility among geometry's properties, making it possible to satisfy all of them by the same parameter. This parameter, in turn, consistently describes all geometrical features of the manifold and thus must be the one used to describe space-time in accordance with our postulate for freely falling test particles.

\section{Variational Formulation for Gravitation in Weyl} \label{sec: variational formulation}
After establishing a geometrical description for space-time in which free test particles dynamics are given by Eq.~(\ref{geodesic equation for proper time}), we now proceed with a theory for space-time itself. Our intent is to describe both dynamics from a variational formulation.

For the space-time lagrangian, we will take the Palatini approach which considers the connection coefficients as independent dynamical variables as well. Since we are considering a Weyl geometry, we end up with three dynamically independent geometrical objects: the metric, the vector $\omega_\mu$ and the connection. However, a satisfactory Palatini formulation that results in the most general case of a Weyl geometry is still not known. What is known, though, is such formulation for the particular case of WIST, where $\omega_\mu = \partial_\mu \omega$. The same happens for the free test particle lagrangian. Additionally, according to Ref.~\citen{Jurgen}, only through WIST it is possible to recover the Hamilton-Jacobi formalism. Therefore, our lagrangian formulation of gravitation will be restricted to WIST and the vector $\omega_\mu$ is replaced by the scalar $\omega$ in the set of dynamical variables.

We will begin with a space-time action for vacuum given by:

\begin{equation}
S_g = \frac{1}{2\kappa}\int e^{-\omega}R\sqrt{-g}d^4x \; ,
\label{vacuum action}
\end{equation}

\noindent where $\kappa = 8\pi G/c^4$, $G$ is Newton's gravitational constant, $c$ is the speed of light, $g$ is the metric determinant and $R = g^{\mu\nu}R_{\mu\nu}$ is the \emph{Ricci} (or \emph{curvature}) \emph{scalar}, given in terms of the \emph{Ricci tensor} $R_{\mu\nu} = R^\alpha_{\mu\alpha\nu}$, in which $R^\alpha_{\beta\gamma\delta}$ is the \emph{Riemann tensor}, given by:

\begin{equation}
R^{\alpha}_{\mu\beta\nu} = \partial_\nu\Gamma^{\alpha}_{\mu\beta} - \partial_\beta\Gamma^{\alpha}_{\mu\nu} + \Gamma^{\alpha}_{\lambda\nu}\Gamma^{\lambda}_{\mu\beta} - \Gamma^{\alpha}_{\lambda\beta}\Gamma^{\lambda}_{\mu\nu} \; .
\label{Riemann tensor}
\end{equation}


Imposing that variations of the action (\ref{vacuum action}) with respect to the connection coefficients should vanish gives the desired WIST connection, which is just (\ref{Weyl connection}) with $\omega_\mu = \partial_\mu\omega$.


Now, if we impose the same for variations of the metric coefficients we obtain:

\begin{equation}
G_{\mu\nu} \equiv R_{\mu\nu} - \frac{1}{2}g_{\mu\nu}R = 0 \; ,
\label{Einstein equation for vacuum}
\end{equation}

\noindent where $G_{\mu\nu}$ is the \emph{Einstein tensor}.

Finally, doing the same for $\omega$ we get $R = 0$, which is precisely the trace of (\ref{Einstein equation for vacuum}) and is thus redundant. This means there is an arbitrariness in the expression for one of the functions involved. The reason for this is the already mentioned gauge invariance of the geometry, which is preserved in this lagrangian formulation and, therefore, allows for an arbitrary gauge function.



To obtain the gauge-invariant Eq.~(\ref{geodesic equation for proper time}) for a free test particle of mass $m$, we proceed the same way as in GR and take the proper-time between two events as its action. After all, in consistency with the local validity of Special Relativity, test particles trajectories should maximize their proper-time and, then, the action is taken to be:

\begin{equation}
S_p = 2m \int d\tau \; .
\label{free test particle action}
\end{equation}

The expression for $d\tau$ is obtained from (\ref{proper-time equation}). Since we can add an arbitrary constant to $\omega$, we can set it conveniently to integrate this equation as:

\begin{align}
& u^2 = g_{\mu\nu}\frac{dx^\mu}{d\tau}\frac{dx^\nu}{d\tau} = e^\omega \label{proper-time integration in WIST} \\
\therefore \qquad & d\tau^2 = e^{-\omega}g_{\mu\nu}dx^\mu dx^\nu \; . \label{proper-time in WIST}
\end{align}

Test particles follow trajectories in an already given geometry, therefore both $\omega$ and $g_{\mu\nu}$ are fixed in (\ref{free test particle action}). Variation of it with respect to the trajectory gives the desired Eq.~(\ref{geodesic equation for proper time}).

From (\ref{proper-time in WIST}) we can see that observer's proper-time in WIST is actually different from the riemannian case, and that space-time is now described by the \emph{effective metric}:

\begin{equation}
\tilde{g}_{\mu\nu} \equiv e^{-\omega}g_{\mu\nu} \; .
\end{equation}

\noindent However, now it is possible to easily see why this is so.

Considering $\omega_\mu = \partial_\mu\omega$ in (\ref{gauge transformation for general Weyl}), one can perfectly well rewrite the gauge transformation in WIST as:

\begin{equation}
\left\{\begin{array}{rcl}
g_{\mu\nu} & \rightarrow & \bar{g}_{\mu\nu} = e^{\Lambda}g_{\mu\nu} \; , \\
\omega & \rightarrow & \bar{\omega} = \omega + \Lambda \; .
\end{array}
\right.
\label{gauge transformation for WIST}
\end{equation}

\noindent This is still a gauge transformation, with an arbitrary $\Lambda = \Lambda(x^\mu)$, for the geometry and proper-time. If we now chose $\Lambda = -\omega$ (\emph{Einstein gauge}), we would get $\bar{\omega} = 0$, bringing us back to Riemann geometry. Since this kind of geometry is just a gauge in WIST, they should be completely equivalent. In fact, after setting this specific expression for $\Lambda$, we get the equivalent riemannian metric\footnote{It is good to note the subtle formal difference between this expression and the effective metric. While, depending on the gauge, $\omega$ and $g_{\mu\nu}$ may have different values and still provide the same $\tilde{g}_{\mu\nu}$, only in the Einstein gauge we have the \emph{metric} $\bar{g}_{\mu\nu} = \tilde{g}_{\mu\nu}$.} $\bar{g}_{\mu\nu} = e^{-\omega}g_{\mu\nu}$ and the interval (\ref{interval definition}), which is also the proper-time in Riemann, would become (\ref{proper-time in WIST}), also justifying this gauge-invariant expression.

\subsection{Coupling with matter fields} \label{subsec: coupling}
Our theory possesses three geometric dynamical fields and we are now with the task of establishing how other physical fields couple with them. Many different couplings have already been proposed \cite{Novello1,Novello2,Novello3,Novello4,Salim-Sautu1,Salim-Sautu2,Salim-Sautu3,Salim-Sautu4,Salim-Sautu5,NST1,NST2}, but we will take the same found in \cite{Weyl-Romero-Friedmann-artigo,Weyl-Romero-arXiv,Poulis-Seminar,Poulis-PhD} for this purpose.

The reason for this choice is easily seen if one realizes that, in consistency with the Einstein gauge, everything works exactly as if the metric in GR was replaced by the effective metric. This is true for the connection (\ref{Weyl connection}) and the actions (\ref{vacuum action}) and (\ref{free test particle action}). Therefore, we will proceed with that same recipe with any other action from GR, \textit{i.e.}, we take any other lagrangian of the form ${\cal L}\left(g^{\mu\nu}, ...\right)\sqrt{-g}$, in GR, and write it as ${\cal L}\left(\tilde{g}^{\mu\nu}, ...\right)\sqrt{-\tilde{g}}$. With this choice, we end up with a completely gauge-invariant theory of gravitation.

At this point, it is important to clarify the difference between this theory and a conformal transformation in GR, as some may think it is. First of all, the later is far from being the whole point of this paper, since it is very clear from the beginning that our proposal is to consider Weyl geometry as a more general and consistent one to describe space-time, in accordance with light-clocks measurements and the properties of light propagation and freely falling test particles \cite{EPS,Woodhouse}. Secondly, the conformal factor that appears in the theory comes from two more fundamental facts: (a) the space-time interval measurement procedure, which relies on the observers' proper-time \cite{Blokhintsev,Graves} and (b) its characterization in accordance with space-time geometric properties and the postulate for freely falling test particles (Eqs.~(\ref{geodesic equation for proper time}), (\ref{modulus variation}) and (\ref{proper-time equation})). It is only because of these two facts together that one realizes the effective metric plays the same role as the metric in GR,\footnote{This is such a subtle issue that many authors still considers the metric as describing space-time even when working in WIST.} and that is why $\tilde{g}_{\mu\nu}$ was given this name.

This equivalence, however, occurs only at the physical level, when measurements are involved; from the geometric point of view they are very different (see (\ref{Weyl metricity})). A Weyl geometry considers actually \emph{two} independent geometrical fields, whereas in Riemann there is only one (the metric) which is not increased by another one after a conformal transformation. In our theory, the metric has not suffered any conformal transformation to become $\tilde{g}_{\mu\nu}$; this is instead a tensor composed of \emph{both} the manifold metric and $\omega$.

Additionally, the full similarity came only after the specific coupling chosen, which could be totally different. The possibility of treating $\omega$ and $g_{\mu\nu}$ independently, not only dynamically, but when establishing the coupling as well, only makes sense in our framework and would not resemble a conformal transformation in GR in a general case. Our choice, nevertheless, was simply based on the same reasoning of GR.

Furthermore, although we will not work on this here, we can consider a new lagrangian term for $\omega$ due to its dynamical independence. This would have no correspondence in GR even after a conformal transformation, since this would not increase the number of dynamical variables.

Our main concern was to be theoretically consistent and the resemblance of the space-time description to a conformal transformation of the metric in GR was only a fluke.

After those remarks, let us now proceed with the full field equations. The gravitational action in the presence of any other physical field reads:

\begin{equation}
S = \int \left[\frac{1}{2\kappa}R(\tilde{g}^{\mu\nu},\Gamma^\alpha_{\beta\gamma}) + {\cal L}(\tilde{g}^{\mu\nu}, ...)\right]\sqrt{-\tilde{g}}d^4x \; ,
\label{gravitational action with source}
\end{equation}

\noindent where $R(\tilde{g}^{\mu\nu},\Gamma^\alpha_{\beta\gamma}) = \tilde{g}^{\mu\nu}R_{\mu\nu}(\Gamma^\alpha_{\beta\gamma})$ and $\tilde{g}^{\mu\alpha}\tilde{g}_{\alpha\nu} = \delta^\mu_\nu$.

Since $\mathcal{L}$ does not depend on the connection, variation of the action with respect to it gives us WIST. Varying the metric will give:

\begin{gather}
G_{\mu\nu}(\tilde{g}^{\mu\nu}) = -\kappa \, T_{\mu\nu}(\tilde{g}^{\mu\nu}, ...) \; , \label{Einstein equation with source} \\
T_{\mu\nu}(\tilde{g}^{\mu\nu}, ...) \equiv \frac{1}{2\sqrt{-\tilde{g}}}\frac{\delta \left[{\cal L}\left(\tilde{g}^{\mu\nu}, ...\right)
\sqrt{-\tilde{g}}\right]}{\delta \tilde{g}^{\alpha\beta}} \; . \label{Energy-momentum tensor in WIST}
\end{gather}

\noindent The left hand side of Eq.~(\ref{Einstein equation with source}) is precisely the Einstein tensor that appears in (\ref{Einstein equation for vacuum}). It was written that way just to emphasize, as it can be easily checked, that it has the same dependence on $\tilde{g}_{\mu\nu}$ that the corresponding tensor in GR has on $g_{\mu\nu}$. Since the same occurs by construction with the lagrangian density in (\ref{Energy-momentum tensor in WIST}) it will also happen with the \emph{energy-momentum tensor} $T_{\mu\nu}$. Therefore, the field equation coming from the variation of the metric in our theory is just the same obtained in GR by the same variation but replacing $g_{\mu\nu}$ with $\tilde{g}_{\mu\nu}$, as it was already expected.

Variation with respect to $\omega$ gives the trace of (\ref{Einstein equation with source}) which is, once again, redundant. Since the gauge symmetry has not been broken by the coupling, we still have solutions up to an arbitrary function, which justifies this redundancy.

Those are the field equations of our gauge-invariant theory of gravitation in WIST. It considers three geometric dynamical variables $(\omega, g_{\mu\nu}, \Gamma^\alpha_{\beta\gamma})$ where, by solving the equations, the gauge-dependent pair $(\omega, g_{\mu\nu})$ provides a gauge-invariant connection. However, space-time is now described by the gauge-invariant effective metric. GR is only a gauge inside this theory.

\section{Static and Spherically Symmetric Vacuum} \label{sec: vacuum solution}
Now that we have our theory properly set, let us consider the simple case of a static vacuum solution with spherical symmetry. This will be very helpful to show the importance of having a proper-time consistently defined the way we did.

Since we are considering a static and spherically symmetric geometry, we will consider only radial dependence. Let $\omega = \omega(r)$ and the metric be given by the interval:

\begin{equation}
ds^2 = A(r)c^2dt^2 - \frac{dr^2}{B(r)} - r^2d\theta^2 - r^2\sin^2\theta d\varphi^2 \; ,
\label{spherically symmetric interval}
\end{equation}


\noindent where $A(r)$, $B(r)$ and $\omega(r)$ are the functions to be determined by the field equations. However, since these are redundant, we already know that one of the functions will be arbitrary, as we will see.

From the field equations, only two of them are independent and we chose the following pair which comes from the $\mu = \nu = 0$ and $\mu = \nu = 1$ components of (\ref{Einstein equation for vacuum}):

\begin{empheq}[left=\empheqlbrace]{align}
& \frac{B'}{rB} - \frac{1}{r^2B} + \frac{1}{r^2} = \frac{1}{2}\frac{B'}{B}\omega' + \frac{2\omega'}{r} + \omega'' - \frac{\left(\omega'\right)^2}{4} \; , \label{Einstein eq 00} \\
& \frac{A'}{rA} - \frac{1}{r^2B} + \frac{1}{r^2} = \frac{1}{2}\frac{A'}{A}\omega' + \frac{2\omega'}{r} - \frac{3\left(\omega'\right)^2}{4} \; , \label{Einstein eq 11}
\end{empheq}

\noindent where the prime means derivative with respect to $r$. This system can be solved analytically with the solution \cite{Poulis-PhD}:

\begin{empheq}[left=\empheqlbrace]{align}
\omega(r) = & \; \frac{2}{3}\ln\left[r\beta^2(r)\right] + \ln C_2 \; , \label{omega solution} \\
A(r) = & \; C_1C_2\left[r\beta^2(r)\right]^{\frac{2}{3}} - 2\beta^2(r) \; , \label{A solution} \\
B(r) = & \; \frac{9\beta^4(r)\left\{C_1C_2\left[r\beta^2(r)\right]^{\frac{2}{3}} - 2\beta^2(r)\right\}}{C_1C_2\left[r\beta^2(r)\right]^{\frac{2}{3}} \left\{r\left[\beta^2(r)\right]' - 2\beta^2(r)\right\}^2} \; , \label{B solution}
\end{empheq}

\noindent where $C_1$ and $C_2$ are arbitrary constants and $\beta(r)$ is an arbitrary function of $r$. We see that one of the three functions on the left hand side is actually arbitrary.

Before we proceed, it should be noticed that, although all the functions involved will be fixed once any one of them is specified, this would be a restriction with no theoretical support, since the dynamical equations do not specify $\beta(r)$. The most general solution is indeed arbitrary, due to the arbitrariness of this function. This feature of the solution was already expected and simply reflects the gauge freedom the theory possesses; specifying $\beta(r)$ is the same as specifying the gauge.

This arbitrariness, in turn, precludes one of having any interpretation of the coordinates by associating them to an observer only based on the interval (\ref{spherically symmetric interval}). As it was already pointed out, the interval is completely meaningless and one should consider (\ref{proper-time in WIST}) for this purpose instead. If we insisted on the metric to describe space-time we would get inconsistent and meaningless results.

Let us naively consider those coordinates in (\ref{spherically symmetric interval}) as describing space-time for an observer at infinity and regard time derivatives as $\dot{x}^\mu = dx^\mu/dt$ when solving the geodesic Eq.~(\ref{geodesic equation for proper time}). If we take the geodesic to be a circular orbit in the equatorial plane, \textit{i.e.}, $\ddot{r} = 0$, $\dot{r} = 0$, $\dot{\theta} = 0$ and $\theta = \pi/2$, we would obtain for the orbital velocity ($v \equiv r\dot{\varphi}$) that $v^2/c^2 = \beta^2(r)$.

This result is completely inconsistent with the very meaning of a gauge freedom, for it identifies a physical observable with the gauge function whereas it should be totaly independent of it. That is, for a gauge invariance one understands the possibility of choosing any gauge function without changing any possible observation, and this is certainly not the case.

The reason for this comes from the erroneous interpretation of $t$ and $r$ as describing time and distance from origin, respectively. This interpretation can only be provided by the effective metric or (\ref{proper-time in WIST}). Now, if we take $d\tau^2$ there still is the arbitrary $\beta(r)$ preventing us from making any possible interpretation of the coordinates. However, if we make the transformations:

\begin{empheq}[left=\empheqlbrace]{align}
& \bar{r} = e^{-\frac{\omega}{2}}r \; , \label{transf for r} \\
& \bar{t} = C_1^{\frac{1}{2}}t \; , \label{transf for t} \\
& C_1 C_2^{\frac{3}{2}} = \frac{c^2}{GM} \; , \label{redef of C2}
\end{empheq}

\noindent then we have:

\begin{equation}
d\tau^2 = \left(1 - \frac{2GM}{c^2 \bar{r}}\right)c^2 d\bar{t}^2 - \frac{d\bar{r}^2}{\left(1 - \frac{2GM}{c^2 \bar{r}}\right)} - \bar{r}^2d\theta^2 - \bar{r}^2\sin^2\theta d\varphi^2 \; ,
\label{Schwarzschild metric for the effective metric}
\end{equation}


\noindent which is precisely the Schwarzschild solution for the effective metric and there is no arbitrariness at all. The dependence on $\beta(r)$ has gone after transformation (\ref{transf for r}) and we were left with an expression which provides an easy interpretation of those coordinates as being time ($\bar{t}$), radial coordinate ($\bar{r}$), polar angle ($\theta$) and azimuthal angle ($\varphi$) for an observer at infinity.

Once again, we see that such interpretation would not be possible from the interval even after those transformations, for $ds^2 = e^\omega d\tau^2$ and there would still be the arbitrary $\beta(r)$ in the expression for $\omega$, keeping the interval physically meaningless.

If we now regard time derivatives as $\dot{x}^\mu = dx^\mu/d\bar{t}$ when solving the geodesic Eq.~(\ref{geodesic equation for proper time}) and consider a circular orbit in the equatorial plane (now with $\ddot{\bar{r}} = 0$, $\dot{\bar{r}} = 0$, $\dot{\theta} = 0$ and $\theta = \pi/2$), we get for the orbital velocity ($v \equiv \bar{r}\dot{\varphi}$) that $v^2 = GM/\bar{r}$. This gauge independent result is just the same from GR, as it should be, since it corresponds to a specific gauge in our theory.

If we take, in particular, the Einstein gauge by imposing $\omega = 0$ in (\ref{omega solution}), this implies that $\beta^2(r) \propto r^{-1}$. Recalling that our previous (erroneous) result for $v = rd\varphi/dt$ was $v^2 = c^2\beta^2$, we see that now it gives a radial dependence in agreement with the corresponding GR prediction, in consistency with this specific gauge. Performing redefinition (\ref{redef of C2}) and time rescaling (\ref{transf for t}), we get precisely $v^2 = c^2\beta^2 = GM/r$ and the same Schwarzschild solution for the metric, which is now equal to the effective metric and from which we can conclude that $(t, r, \theta, \varphi)$ actually describes space-time for an observer at infinity. Notice also that $\bar{r} = r$ in this gauge, therefore this result agrees with that obtained for the more general one: $v^2 = GM/\bar{r}^2 = GM/r^2$.

From this example, it should be clear that one must refer to the effective metric in order to have a meaningful description of space-time. The whole theory is gauge-invariant, as can be seen from the actions (\ref{free test particle action}) and (\ref{gravitational action with source}), so there cannot be any gauge-dependent measurement.\footnote{Notice this assertion only holds because of the specific gauge-invariant coupling with physical fields we established. Nevertheless, in vacuum this is always true.} When attempting to provide a consistent description, this turns out to be also consistent with the more fundamental idea of ascribing to the geometry of space-time all the gravitational effects on test particles. In other words, a gauge-invariant description is also consistent with Eq.~(\ref{geodesic equation for proper time}).

Regarding the geodesic equation, recall that it gives the same curve irrespective of the parameter used. The key point is that the one which provides the postulated form of Eq.~(\ref{geodesic equation for proper time}) must be the same one considered for space-time description and for the free test particle action (\ref{free test particle action}). That is, all of them refer to the same parameter which we call proper-time: space-time is described by the same parameter which is a maximum between two points of a free test particle trajectory.

Had we insisted that proper-time was given by the interval, then space-time would be described by the gauge-dependent metric and test particles trajectories would be accelerated curves in disagreement with the idea expressed by (\ref{geodesic equation for proper time}). This is not actually a big problem, just a different postulate, however, since the field equations would still be gauge-invariant, they would not suffice to determine the geometry of space-time, forcing us to break the gauge symmetry somehow. If we changed the coupling with physical fields for that purpose, the problem would remain for the vacuum. Therefore, we would have to introduce a new gauge-dependent lagrangian term in action (\ref{vacuum action}). We will not be concerned with such situation in this paper, but it will rather be subject of a future work.

\section{Gauge-Invariant Thermodynamics} \label{sec: thermodynamics}
We now proceed with the description of source fields according to their thermodynamical regime in this gauge-invariant context.\footnote{This will be an extended, and more detailed version of our previous work, Ref.~\citen{Poulis-IGAC}.} For that purpose, we will perform a decomposition of $T_{\mu\nu}$ in terms of their components along the fluid velocity $u^\mu$ and perpendicular to it.

Let

\begin{equation}
h^\alpha_\beta \equiv \delta^\alpha_\beta - (u^\gamma u_\gamma)^{-1}u^\alpha u_\beta = \delta^\alpha_\beta - e^{-\omega}u^\alpha u_\beta
\label{projector}
\end{equation}

\noindent be a (gauge-invariant) projector onto the three-space orthogonal to $u^\mu$, where we have used (\ref{proper-time integration in WIST}). Taking $\kappa = c = 1$ from now on, we then define

\begin{gather}
\rho \equiv T_{\mu\nu}u^\mu u^\nu \; , \label{energy density} \\
P \equiv -\frac{1}{3}e^\omega T_{\mu\nu}h^{\mu\nu} \; , \label{total pressure} \\
q_\mu \equiv h^\alpha_\mu u^\beta T_{\alpha\beta} \; , \label{heat-flow} \\
\pi_{\mu\nu} \equiv h^\alpha_\mu h^\beta_\nu T_{\alpha\beta} - \frac{1}{3}h_{\mu\nu}h^{\alpha\beta}T_{\alpha\beta} \; . \label{anisotropic pressure}
\end{gather}

\noindent as being the \emph{energy density}, \emph{total pressure}, \emph{heat-flow} and \emph{anisotropic pressure}, respectively. Since all of them are gauge-invariant, we see from the Einstein gauge they actually have the same interpretation as in GR. With those definitions, the energy-momentum tensor can be identically rewritten as

\begin{equation}
T_{\mu\nu} = \rho e^{-2\omega}u_\mu u_\nu - Pe^{-\omega}h_{\mu\nu} + e^{-\omega}\left(q_{\mu}u_{\nu} + q_{\nu}u_{\mu}\right) + \pi_{\mu\nu} \; .
\label{energy-momentum tensor decomposition}
\end{equation}

Next, we consider a splitting of this tensor in equilibrium ($\bar{T}_{\mu\nu}$) and dissipative ($\Delta T_{\mu\nu}$) parts. The same goes for $P$, which is split into an \emph{isotropic} ($p$) and \emph{bulk} ($-\pi$) \emph{pressure}, corresponding to the equilibrium and dissipative cases, respectively:

\begin{equation}
P = p - \pi \; .
\label{splitting of total pressure}
\end{equation}

Since it is supposed that both of them may occur independently, then both should be gauge-invariant like $P$.

Now we write

\begin{gather}
\Delta T_{\mu\nu} \equiv e^{-\omega}\left(q_{\mu}u_{\nu} + q_{\nu}u_{\mu}\right) + \pi e^{-\omega}h_{\mu\nu} + \pi_{\mu\nu} \; , \label{dissipative part of EM tensor} \\
\bar{T}_{\mu\nu} \equiv \rho e^{-2\omega}u_\mu u_\nu - pe^{-\omega}h_{\mu\nu} \; , \label{equilibrium part of EM tensor} \\
\therefore \quad T_{\mu\nu} = \bar{T}_{\mu\nu} + \Delta T_{\mu\nu} \; . \label{EM tensor splitting}
\end{gather}

In the following, we will adopt Eckart's frame \cite{Eckart}, where $u^\mu$ corresponds to the velocity of the fluid particles, from which we have the particle flux:

\begin{equation}
N^\mu = nu^\mu \; ,
\label{particle flux}
\end{equation}

\noindent where $n$ is the number of particles per unit volume and is taken to be gauge-invariant.

Before we proceed, it is important to notice that Bianchi identity still holds in Weyl, even in the most general case:

\begin{equation}
\nabla_\mu R^\alpha_{\beta\gamma\lambda} + \nabla_\lambda R^\alpha_{\beta\mu\gamma} + \nabla_\gamma R^\alpha_{\beta\lambda\mu} = 0 \; .
\label{Bianchi identity}
\end{equation}

\noindent When one considers WIST, this implies

\begin{equation}
\nabla_\alpha\left(e^{2\omega}G^{\alpha\gamma}\right) = 0 \quad \therefore \quad \nabla_\alpha\left(e^{2\omega}T^{\alpha\gamma}\right) = 0 \; , \label{Einstein and EM tensor divergence}
\end{equation}


\noindent where the second equation comes by virtue of (\ref{Einstein equation with source}).

Inserting decompositions (\ref{equilibrium part of EM tensor}) and (\ref{EM tensor splitting}) into this result, one obtains:

\begin{gather}
\dot{\rho} + \left(\rho + p\right)\theta = -e^{-\omega}u_\mu\nabla_\nu\left(e^{2\omega}\Delta T^{\mu\nu}\right) \; ,
\label{energy equilibrium}
\end{gather}


\noindent where $\theta \equiv \nabla_\mu u^\mu $ and the dot is now regarded as the application of $u^\mu\nabla_\mu$ on the corresponding quantity ($\dot{\rho} = u^\mu\nabla_\mu\rho$). The quantity $\theta$ is gauge-invariant and, therefore, has the same interpretation as its riemannian equivalent, giving the rate of expansion of a three-volume orthogonal to $u^\mu$.

We now take the following equations:

\begin{gather}
Tds = \frac{1}{n}d\rho + (\rho + p)d\left(\frac{1}{n}\right) \label{Gibbs equation} \\
\therefore \quad T\dot{s} = \frac{1}{n}\dot{\rho} - (\rho + p)\frac{\dot{n}}{n^2} \; , \label{time derivative of entropy} \\
Ts = \frac{1}{n}\left(\rho + p\right) - \mu \; , \label{chemical potential}
\end{gather}

\noindent where the first is the Gibbs equation and the last introduces the \emph{chemical potential},\cite{Israel} $\mu$. There we also have the \emph{entropy per particle}, $s = s(\frac{\rho}{n},\frac{1}{n})$, which can be seen from its dependence that it is gauge-invariant and, therefore, both the \emph{temperature}, $T$, and $\mu$ are also gauge-invariant.

Next, we define the following quantities:

\begin{gather}
\beta_\mu \equiv \frac{e^{-\omega}}{T}u_\mu \; , \label{constant entropy Killing vector} \\
\psi \equiv \nabla_\mu N^\mu = \dot{n} + n\theta \; , \label{particle creation rate} \\
s^\mu \equiv nsu^\mu + \frac{e^{\omega}}{T}u_\nu\Delta T^{\mu\nu} \; , \label{entropy flux} \\
V_{\alpha\beta} \equiv h^\lambda_\alpha h^\delta_\beta \nabla_\delta u_\lambda = \omega_{\alpha\beta} + \sigma_{\alpha\beta} + \frac{1}{3}\theta h_{\alpha\beta} \; , \label{rotation, shear and expansion}
\end{gather}

\noindent Where $\psi$ is the \emph{particle creation rate} and $s^\mu$ is the \emph{entropy current vector} \cite{Weinberg}. The last equality, in turn, is also an identity, since we have:

\begin{gather}
\theta_{\alpha\beta} \equiv V_{(\alpha\beta)} \quad \Rightarrow \quad \theta = {\theta^\alpha}_\alpha \; , \label{expansion} \\
\sigma_{\alpha\beta} \equiv \theta_{\alpha\beta} - \frac{1}{3}\theta h_{\alpha\beta} \; , \label{shear} \\
\omega_{\alpha\beta} \equiv V_{[\alpha\beta]} \; . \label{rotation}
\end{gather}

\noindent In those expressions, and whenever they appear, the parenthesis and square brackets denote symmetrization and anti-symmetrization, respectively, on the corresponding pair of indices, so that $V_{\alpha\beta} = V_{(\alpha\beta)} + V_{[\alpha\beta]}$. Once again, since ${\sigma^\alpha}_\beta$ and ${\omega^\alpha}_\beta$ are gauge-invariant, they also have the same interpretation as their riemannian equivalent, giving the shear and rigid rotation of a three-volume orthogonal to $u^\mu$, respectively.

From (\ref{projector}) and (\ref{rotation, shear and expansion}), we can write:

\begin{equation}
\nabla_\nu u_\mu = \sigma_{\nu\mu} + \frac{1}{3}\theta h_{\nu\mu} + \omega_{\mu\nu} + e^{-\omega}u_\nu\dot{u}_\mu + u_\mu\omega_\nu - e^{-\omega}u_\nu u_\mu u^\gamma\omega_\gamma \; .
\label{velocity divergence in terms of rotation, shear and expansion}
\end{equation}


\noindent Finally, from Eqs.~(\ref{dissipative part of EM tensor}), (\ref{energy equilibrium}), (\ref{time derivative of entropy})-(\ref{entropy flux}) and (\ref{velocity divergence in terms of rotation, shear and expansion}) we have the \emph{entropy balance}:

\begin{gather}
\nabla_\mu s^\mu + \frac{\psi\mu}{T} = e^{2\omega}\Delta T^{\mu\nu}\nabla_\nu\beta_\mu = \label{entropy divergence due to beta} \\
= e^\omega\left[\frac{\sigma_{\mu\nu}\pi^{\mu\nu}}{T} + \frac{\theta\pi e^{-\omega}}{T} - \left(\partial_\mu T - e^{-\omega}T\dot{u}_\mu\right)\frac{q^\mu}{T^2}\right] \; . \label{entropy divergence}
\end{gather}

\noindent The difference between the incoming and outgoing entropy from an infinitesimal volume ($\nabla_\mu s^\mu$) is related to a source of entropy (other terms) through this equation.

From now on we will take $\psi = 0$. Therefore, in order to ensure a positive entropy variation, we should have \cite{Weinberg}:

\begin{empheq}[left=\empheqlbrace]{align}
\pi_{\mu\nu} & = \eta e^{-\omega}\sigma_{\mu\nu} \; , \label{shear viscosity} \\
\pi & = \zeta\theta \; , \label{bulk viscosity} \\
q_\mu & = \chi h^\alpha_\mu\left(\partial_\alpha T - e^{-\omega}T\dot{u}_\alpha\right) \; , \label{heat conduction}
\end{empheq}

\noindent where the gauge-invariant quantities $\eta$, $\zeta$ and $\chi$ are the \emph{shear viscosity}, \emph{bulk viscosity} and \emph{heat conduction}, respectively, and none of them is negative. Eq.~(\ref{entropy divergence}) then reads ($\psi = 0$):

\begin{equation}
\nabla_\mu s^\mu = e^{2\omega}\frac{\pi_{\mu\nu}\pi^{\mu\nu}}{\eta T} + \frac{\pi^2}{\zeta T} - e^\omega\frac{q_\mu q^\mu}{\chi T^2} \; .
\label{entropy flux positive definite}
\end{equation}

Before we proceed, let us note that if $\beta_\mu$ was a Killing vector, we would have from (\ref{entropy divergence due to beta}):

\begin{equation}
\nabla_{\left(\nu\right.} \beta_{\left.\mu\right)} = 0 \quad \Rightarrow \quad \nabla_\mu s^\mu = 0 \; ,
\label{constant entropy}
\end{equation}

\noindent since we are considering $\psi = 0$.

\section{Geometric Scalar Field and Physical Content} \label{sec: examples}
We have already made clear the equivalence between $\tilde{g}_{\mu\nu}$ in our theory and $g_{\mu\nu}$ in GR, and that we are not by any means simply performing a conformal transformation in the later, but rather starting from first principles.

In this sense, two geometries that differ only in their expressions for $\omega$, \textit{i.e.}, one described by $\tilde{g}^A_{\mu\nu} = e^{-\omega_A}g_{\mu\nu}$ and the other by $\tilde{g}^B_{\mu\nu} = e^{-\omega_B}g_{\mu\nu}$ will, in general, be associated with different physical situations. This is quite reasonable, since $\tilde{g}^A_{\mu\nu} = F\tilde{g}^B_{\mu\nu}$, with $F = e^{-\omega_A}e^{\omega_B}$ and, hence, they are as much different as two conformally related geometries in GR. Therefore, although the specific role played by each of the geometrical fields considered is rather vague, due to the gauge invariance, it is clear that both of them have a certain physical significance, in the sense that geometries that differ only in $\omega$ or only in $g_{\mu\nu}$ do not correspond, in general, to the same physical situation.

Concerning conformally flat space-times, which are characterized by a null Weyl tensor\footnote{The Weyl tensor is given by \[C_{\alpha\beta\gamma\delta} \equiv R_{\alpha\beta\gamma\delta} - R_{\alpha\left[\gamma\right.}g_{\beta\left.\delta\right]} + R_{\beta\left[\gamma\right.}g_{\alpha\left.\delta\right]} + \frac{1}{3}Rg_{\alpha\left[\gamma\right.}g_{\beta\left.\delta\right]} \; .\]} in the Einstein gauge, the relation between $\omega$ and the physical content is even more notorious. In Ref.~\citen{Poulis-Seminar} we pointed out that one can always write $g_{\mu\nu} = e^{-\Lambda}\eta_{\mu\nu}$ for this kind of geometry and it was explicitly shown for the case of cosmological models, where $\Lambda = \Lambda(x^\mu)$ and $\eta_{\mu\nu}$ is the Minkowski metric. If, moreover, we had such geometry in the Einstein gauge, one could then perform transformation (\ref{gauge transformation for WIST}), with this same $\Lambda$ as the gauge function, and the geometry would depart from a null $\omega$ and varying $g_{\mu\nu}$ to an equivalent one with varying $\bar{\omega} = \Lambda$ and fixed $\bar{g}_{\mu\nu} = \eta_{\mu\nu}$.

As it was extensively emphasized here in this paper, this equivalence is naturally expected, as a result of the gauge invariance. Nevertheless, it was explicitly verified in Refs.~\citen{Poulis-Seminar,Poulis-PhD} for the red-shift, which is one of the most important observables in cosmology, for example.

Therefore, in the case of conformally flat space-times, the geometric scalar field $\omega$ can incorporate all geometrical features of the space-time.\footnote{Apart from the fixed $g_{\mu\nu} = \eta_{\mu\nu}$.} Since those are intimately related to the physical content considered, one can clearly see that different expressions for $\omega$ (for the same metric) actually correspond to different physical systems.

Curiously, despite that correspondence, nothing changes with regard to light propagation. Although the underlying geometry may be associated with different physical situations, light rays trajectories are all the same irrespective of the expression for $\omega$, as long as the metric is always the same \cite{Poulis-Seminar}. In fact, even in the most general case of Weyl geometries, where $\omega_\mu$ is not necessarily a divergent of a scalar, any null geodesic is insensitive to the specific value of this vector field; all that matters is the expression for the metric\footnote{This fact may indeed be considered as an evidence for the need of considering Weyl geometries when describing space-time by the use of a measuring device based on light rays propagation.} \cite{Poulis-PhD}.

\subsection{Conformally flat examples}
Once we have realized that a change in the expression for $\omega$, for the same metric, may correspond to a different physical situation, it is interesting to determine how the corresponding thermodynamical regime changes in response to this modification. This is possible now we have the results from the last section and we will give here two examples of this relation between $\omega$ and the physical content together with their thermodynamical description.

We first take a Friedmann-Lema\^{\i}tre-Robertson-Walker (FLRW) model, with plane section for simplicity, in the Einstein gauge ($\omega = 0$),

\begin{align}
ds^2 & = dt^2 - a^2(t)\left(dx^2 + dy^2 + dz^2\right) \label{FLRW metric plane section cosmological time} \\
& = \alpha^2(\eta)\left(d\eta^2 - dx^2 - dy^2 - dz^2\right) \label{FLRW metric plane section conformal time}
\end{align}

\noindent where $\alpha(\eta) = a\circ t(\eta) = a(t(\eta))$ and $d\eta^2 = a^{-2}(t)dt^2$. Then we set a new expression for $\omega$ to make the transition to another space-time configuration associated with a different source. In the first example we depart from a perfect fluid configuration characterized by a \emph{scale factor}, $a(t)$, given by $a(t) = a_0t^N$, and then we get a space-time associated with a new kind of cosmological constant solution. The second example departs from a general scale factor and arrives at a static configuration.

\subsubsection{Cosmological constant from a general perfect fluid}
In a co-moving frame in the Einstein gauge, a perfect fluid in equilibrium is described by the following energy-momentum tensor and equation of state:

\begin{gather}
T_{\mu\nu} = \rho u_\mu u_\nu - ph_{\mu\nu} \; , \quad u^\mu = \delta^\mu_0 \; , \label{perfect fluid EM tensor and velocity} \\
p = \lambda \rho \; , \label{perfect fluid equation of state}
\end{gather}

\noindent where we must have $\lambda < 1$ to prevent sound waves faster than light \cite{Weinberg}, for $\lambda = -1$ we have the so-called \emph{cosmological constant}.

Inserting those expressions together with the metric given by interval (\ref{FLRW metric plane section cosmological time}) in Eq.~(\ref{Einstein equation with source}) and solving for $a(t)$ we have two types of solutions: one for $\lambda \neq -1$ and another for $\lambda = -1$. The first possibility gives

\begin{gather}
a(t) = a_0t^N \; , \quad N \equiv \frac{2}{3(\lambda + 1)} \; , \label{FLRW scale factor for a perfect fluid not CC} \\
\frac{p}{\lambda} = \rho = \frac{3N^2}{t^2} \; , \label{pressure and energy density for perfect fluid not CC}
\end{gather}

\noindent where $a_0$ is an arbitrary constant. In this case, in order to satisfy (\ref{shear viscosity})-(\ref{heat conduction}) in the absence of dissipative terms ($\Delta T_{\mu\nu} = 0$), we must have $\zeta = 0$ and a temperature $T = T(t)$, (no spatial dependence) which is naturally expected by virtue of the implicit homogeneity assumed in (\ref{FLRW metric plane section cosmological time}).

Now, for $\lambda = -1$ we have

\begin{gather}
a(t) = a_0\exp\left(\sqrt{\frac{\Lambda}{3}}t\right) \; , \label{FLRW scale factor for CC} \\
-p = \rho = \Lambda \; , \label{pressure and energy density for CC}
\end{gather}

\noindent where $\Lambda$ is also an arbitrary constant.

Before attempting to establish constraints from (\ref{shear viscosity})-(\ref{heat conduction}), it is interesting to note that (\ref{pressure and energy density for CC}) implies $T_{\mu\nu} = \Lambda g_{\mu\nu}$, whose decomposition (\ref{energy density})-(\ref{anisotropic pressure}) always gives $-p = \rho = \Lambda$ and all the others ($\pi$, $\pi_{\mu\nu}$ and $q_\mu$) equal to zero\footnote{We can consider $P = p \; \Leftrightarrow \; \pi = 0$ with no loss of generality.} regardless the expression for the velocity and metric. Therefore, the right-hand side of Eqs.~(\ref{shear viscosity})-(\ref{heat conduction}) will only be zero for every $u^\mu$ and $g_{\mu\nu}$ if, and only if, we have $\eta$, $\zeta$ and $\xi$, all of them equal to zero.

Now, if we take the metric solution given by (\ref{FLRW scale factor for a perfect fluid not CC}), for $\lambda \neq -1$, and consider $\omega$ not anymore equal to zero, but given by:

\begin{eqnarray}
e^{\omega(t,x)} = B_0\left[a(t)\left(\frac{x}{2} + A(t)\right)\right]^2 \; ,
\label{CC solution}
\end{eqnarray}

\begin{equation}
A(t) = \left\{\begin{array}{lll}
\frac{A_0a^{\left(\frac{1}{N} - 1\right)}}{2a_0^{\frac{1}{N}}\left(1 - N\right)} + A_1 & \quad (N \neq 1) \; , & \\
& & \\
\frac{A_0}{2a_0}\ln\left(A_1 a\right) & \quad (N = 1) \; ,
\end{array}
\right.
\label{CC solution II}
\end{equation}

\noindent where $A_0$, $B_0$ (both positives) and $A_1$ are constants, we make the transition to a cosmological constant solution with

\begin{equation}
-p = \rho = \Lambda = \frac{3}{4}B_0\left(A_0^2 - 1\right) \; .
\label{cosmological constant value}
\end{equation}

However, the proper-time associated with this geometry is given by

\begin{align}
d\tau^2 & = e^{-\omega}\alpha^2(\eta)\left(d\eta^2 - dx^2 - dy^2 - dz^2\right) \nonumber \\
& = \frac{4}{B_0\left(x \pm A_0\eta\right)^2}\left(d\eta^2 - dx^2 - dy^2 - dz^2\right) \; , \label{CC anisotropic proper-time}
\end{align}

\noindent whereas the one associated with the previous cosmological constant solution, given by (\ref{FLRW metric plane section conformal time}) and (\ref{FLRW scale factor for CC}) in the Einstein gauge, is written as

\begin{equation}
d\tau^2 = ds^2 = \frac{3}{\Lambda\eta^2}\left(d\eta^2 - dx^2 - dy^2 - dz^2\right) \; .
\label{CC proper-time}
\end{equation}

In both cases, the zero of $\eta$ was conveniently chosen. Nevertheless, regardless of this choice, there is no way of performing any coordinate transformation in order to put them both in the same form. Since the later corresponds to a homogeneous and isotropic universe, we may infer that this is no longer the case of (\ref{CC anisotropic proper-time}).

Curiously, when computing the following (gauge-invariants) curvature invariants: $e^{\omega}R$, $e^{2\omega}R^{\mu\nu}R_{\mu\nu}$ and $e^{2\omega}R^{\alpha\beta\mu\nu}R_{\alpha\beta\mu\nu}$ for the new solution, we see that all of them are constants; they do not exhibit any coordinate dependence at all.\footnote{Although $R$, $R^{\mu\nu}R_{\mu\nu}$ and $R^{\alpha\beta\mu\nu}R_{\alpha\beta\mu\nu}$ are also curvature invariants, those are not relevant for the observer since their contractions does not involve the effective metric and they are, consequently, gauge-dependent.}

\subsubsection{Static model in thermodynamical equilibrium}
From Eq.~(\ref{FLRW metric plane section conformal time}), the possibility of moving to a drastically different kind of geometry by a simple choice for $\omega$ is notorious.

Starting from this interval in the Einstein gauge, with a general expression for $a(t)$, we change the expression for $\omega$ from zero to:

\begin{equation}
\omega = 2\ln\left[T(x^i)a(t)\right] + \omega_0 \; ,
\label{omega in terms of temperature}
\end{equation}

\noindent where $\omega_0$ is an arbitrary constant and $i \in \{1, 2, 3\}$. With this expression, space-time will no longer be described by a proper-time given by the interval (\ref{FLRW metric plane section conformal time}), but rather by:

\begin{equation}
d\tau^2 = \frac{e^{-\omega_0}}{T^2(x^i)}\left(d\eta^2 - dx^2 - dy^2 - dz^2\right) \; .
\label{static metric in equilibrium}
\end{equation}

If $T(x^i)$ is the fluid temperature and the observer is co-moving with it ($u^\mu = e^{\frac{\omega}{2}}\delta^\mu_0$), then (\ref{constant entropy}) is satisfied and the source is in thermodynamical equilibrium regardless the existence of dissipative terms in the energy-momentum tensor.

On the other hand, when considering Eq.~(\ref{Einstein equation with source}) in (\ref{heat-flow}) we see that $q_\mu$ is identically null. Moreover, the kinematic quantities $\theta$, $\sigma_{\mu\nu}$ and $\omega_{\mu\nu}$ are also null and, in accordance with (\ref{shear viscosity}) and (\ref{bulk viscosity}), we should have $\pi_{\mu\nu}$ and $\pi$ both null as well.

Replacing, now, the energy-momentum tensor by $G_{\mu\nu}$ in (\ref{anisotropic pressure}) and imposing it should be zero, we obtain the following condition for the temperature:

\begin{equation}
T(\mathbf{r}) = k\mathbf{r}^{2} + \mathbf{r}_0\cdot\mathbf{r} + T_0 \; ,
\label{static solution temperature}
\end{equation}

\noindent where $\mathbf{r} = (x,y,z)$, $\mathbf{r}_0$ is an arbitrary constant vector and both $k$ and $T_0$ are arbitrary constants.

Doing the same replacement in (\ref{energy density}) and (\ref{total pressure}), we get:

\begin{gather}
\rho = 3e^{\omega_0}\left(4kT_0 - 3\mathbf{r}^{2}_0\right) \equiv \rho_0 \; , \label{static solution energy density} \\
p = 4e^{\omega_0}kT - \rho_0 \; , \label{static solution pressure}
\end{gather}

\noindent and from (\ref{Gibbs equation}) and (\ref{chemical potential}) we have:

\begin{gather}
s = \frac{\left(\rho_0 + p\right)}{nT} + s_0 \; . \label{static solution entropy} \\
\mu = -s_0 T \; . \label{static solution chemical potential}
\end{gather}

These two examples relate conformally flat geometries\footnote{$C_{\alpha\beta\gamma\delta} = 0$ in the Einstein gauge for all of them: (\ref{FLRW metric plane section conformal time}), (\ref{CC anisotropic proper-time}) and (\ref{static metric in equilibrium}).} and clearly show the expected relation between $\omega$ and the source. One may ask whether they are physically possible or not, however, although both are thermodynamically acceptable, that was not our concern. They are here only to show how different expressions for the geometric scalar field can affect space-time and be associated with different physical systems.

\section{Conclusions and Final Remarks} \label{sec: conclusions}
Following the conclusions of EPS, we performed a reformulation of Einstein's gravitation theory in the broader context of Weyl geometries. A judicious definition of proper-time was given as a consequence of the postulate for freely falling test particles alone, irrespective of the particular underlying geometry and, therefore, contemplating all of them, which also meets the definition given in Ref.~\citen{Perlick}. The close relationship between proper-time and space-time description was pointed out and it was noted the later is not accomplished by the interval in a general case.

When attempting to provide a variational formulation, we were forced to restrict ourselves to the intermediary case of WIST, which is still more general then Riemann geometry but not the most general one as allowed by light-clocks measurements. However, by virtue of the gauge symmetry present in Weyl geometries, both WIST and Riemann become equivalent, since the later is obtained by a simple gauge transformation.

Regarding observers' description of space-time, because of the consistent definition of proper-time we made, it turns out to be gauge-invariant and, therefore, space-time description in WIST is completely equivalent to the riemannian case. The postulate for freely falling test particles is then recovered by imposing that the proper-time between any two events must be a maximum.

After realizing the emergence of the effective metric as the tensor which plays the same role as the metric in GR, our choice for the coupling between geometrical and other physical fields was straightforward; we simply coupled them with the effective metric in the same manner they couple with the metric in GR. Thanks to this coupling we obtained a gauge-invariant gravitation theory equivalent to GR.

Next, we provided a new exact analytical solution of the field equations for a static and spherically symmetric vacuum. This served as an explicit example which shows that space-time actually must be described by the consistent proper-time we established.

To complete our treatment, we reformulated the description of the source fields according to their thermodynamical regime in the context of WIST. From the equivalence between the effective metric in our theory and the metric in GR, this reformulation was straightforward and easily justified by taking the Einstein gauge. Nevertheless, this recourse has a greater appeal in our work due to the previously noticed role of the effective metric.

After that, we gave two examples of how one can depart from a FLRW geometry and arrive at a totally different physical situation only by changing $\omega$. This not only indicates an association of this geometrical field with the physical content, but also reveals this procedure as an useful tool to generate new GR solutions from an already known one; it is just a matter of considering the resulting geometry in the Einstein gauge.

We ended up with a gauge-invariant theory of gravitation provided with two geometrical objects in accordance with EPS conclusions about the geometry of space-time. It recovers General Relativity by a gauge transformation and, consequently, we do not provide any new prediction that could privilege our formulation. Its only distinction from GR has been conceptual, in order to make it theoretically more consistent.

Nevertheless, from the fundamental level of the discussion, we believe our results could be availed in any other situation, specially when it comes to determine the frame (Einstein, Jordan or whatever) one should work on. From the importance assigned to proper-time in describing space-time and its relation with test particles action, one must refer to the later in order to infer the correct frame. Once this has been done, it will correspond to an effective metric and this will be the one that will give information about the space-time geometry. Therefore, whenever there are boundary or symmetry conditions on space-time, they should be reflected on the effective metric.

This point was particularly clear in our treatment mainly because of the gauge symmetry. However, we state that even in the absence of such property the correct space-time description should be taken from test particles lagrangian just like we did.\footnote{We regard the gauge-invariance of our theory as a helpful tool to prove this statement.} In this sense, if we had considered a specific lagrangian term for $\omega$ that could eventually break the gauge symmetry, the effective metric would still play the same role.

This possibility will be considered in a future paper, where it will be shown that the addition of such a lagrangian term would make the theory trivially identified with GR in the presence of a scalar field.

\section*{Acknowledgment}
We acknowledge CNPq, CAPES and FAPEMA for financial support.

\bibliographystyle{ws-ijmpd}
\bibliography{references}

\end{document}